\documentclass[twocolumn,showpacs,preprintnumbers,amsmath,amssymb]{revtex4}

\usepackage[dvips]{graphicx}
\usepackage{dcolumn}
\usepackage{bm}

\begin{document}

\title{High temperature expansion of  emptiness formation probability 
for isotropic Heisenberg chain
} 

\author{Zengo Tsuboi and Masahiro Shiroishi}

\affiliation{%
Institute for Solid State Physics, University of Tokyo, \\
Kashiwanoha 5-1-5, Kashiwa, Chiba, 277-8581 Japan 
}%

\date{}

\begin{abstract}
Recently, G\"ohmann, Kl\"umper and Seel have derived novel integral formulas for the 
correlation functions of the spin-${1/2}$ Heisenberg chain at finite temperature. 
We have found that the high temperature expansion (HTE) technique can be effectively 
applied to evaluate these integral formulas. Actually, as for the emptiness formation 
probability ${P(n)}$ of the isotropic Heisenberg chain, we have found a general formula 
of the HTE for ${P(n)}$ with arbitrary $n \in {\mathbb Z}_{\ge 2}$ up to ${O((J/T)^{4})}$. 
If we fix a magnetic field to a certain value, we can calculate the HTE to much higher
 order. For example, the order up to ${O((J/T)^{42})}$ has been achieved in the case of 
${P(3)}$ when ${h=0}$. We have compared these HTE results with the data by Quantum Monte 
Carlo simulations. They exhibit excellent agreements. 
\end{abstract}

\maketitle

The spin-${1/2}$ Heisenberg chain has been one of the most 
fundamental models in the study of the low dimensional magnetism, partially because 
it can be solved exactly by Bethe ansatz. In fact, many physical quantities of 
the model have been  evaluated {\it exactly} even at finite temperature \cite{TakaBook}. 
However, they are usually the bulk properties, which may be derived directly from the 
free-energy of the system. The exact evaluation of the correlation 
functions at finite temperature, on the other hand, has remained to be a much more difficult problem. 
Actually the evaluation of the correlation functions is still not solved fully even 
in the case of the ground state, although there are various developments recently \cite{JM95,KIEU94,
KMT00,KMT02}. Note, however, that it is established that the correlation functions 
in the ground state are expressed in terms of the multiple integrals. More recent researches try to 
evaluate these integrals, thereby to obtain the precise numerical values of 
the correlation functions \cite{BK01,BKNS02,BKS04,SSNT03,BST04}. 

Considering these situations, it was a great surprise that G\"ohmann, Kl\"umper and Seel 
\cite{GKS04-1,GKS04-2,GKS04-3} succeeded in generalizing the multiple integral formulas to 
the finite temperature, by combining the algebraic Bethe ansatz technique and the quantum transfer 
matrix approach \cite{S85,SI87,K92,DD92,S03}. Their results will be a basis for the further study of 
the model at finite temperature. Then naturally, it is a significant problem to explore these 
integrals and find a way to extract numerical values of the correlation functions at 
finite temperature. Unfortunately it is not a straight forward task to generalize those methods developed 
in the case of the ground state to the finite temperature. Because the latter formula contains 
an additional auxiliary function $\mathfrak{a}(v)$, which is a solution of a certain 
non-linear integral equations (NLIE), and is more complicated. The purpose of this letter is to attack 
this challenging problem with the high temperature expansion (HTE) technique. Surprisingly enough, once 
we introduce the HTE, we can perform the multiple integrals for each term in the HTE series simply by taking 
a residue at the origin.

The Hamiltonian of the spin-${1/2}$ isotropic Heisenberg chain in a 
magnetic field $h$ is defined by
\begin{eqnarray}
H=J\sum_{j=1}^{L}(\sigma_{j}^{x}\sigma_{j+1}^{x}+
\sigma_{j}^{y}\sigma_{j+1}^{y}
+\sigma_{j}^{z}\sigma_{j+1}^{z})-\frac{h}{2}\sum_{j=1}^{L}\sigma_{j}^{z},
\end{eqnarray}
where $\sigma_{j}^{k}$ ($k=x,y,z$) are the Pauli matrices $\sigma^{k}$ 
acting non-trivially on the $j$-th site of the chain of length $L$. Here we adopt the periodic 
boundary condition $\sigma_{j+L}^{k}=\sigma_{j}^{k}$. In this letter, we mainly consider a special 
correlation function called the emptiness formation probability (EFP) ${P(n)}$, which is the probability 
of $n$ adjacent spins being aligned upward :  
\begin{eqnarray}
P(n)=\frac{{\rm Tr}\left\{e^{-\frac{H}{T}}
\prod_{j=1}^{n}\frac{1+\sigma_{j}^{z}}{2}\right\}
}
{{\rm Tr} e^{-\frac{H}{T}}}.
\end{eqnarray}
At zero temperature $T=0$, it was introduced in \cite{KIEU94} and studied further, for example, 
in \cite{BK01,BKNS02,BKS04}. Recently G\"ohmann, Kl\"umper and Seel
 obtained \cite{GKS04-1,GKS04-2} 
the multiple integral formulas of the EFP at finite temperature as 
\begin{eqnarray}
&& P(n)=\left[\prod_{j=1}^{n}\int_{C}\frac{dy_{j}}{2\pi (1+\mathfrak{a}(y_{j}))} \right]
 \label{efp} \\ 
&& \hspace{-10pt} \det_{1\le j,k \le n}
 \left(\frac{\partial_{\xi}^{(k-1)} G(y_{j},\xi)|_{\xi=0}}{(k-1)!} \right)
 \frac{\prod_{j=1}^{n}(y_{j}-i)^{j-1}y_{j}^{n-j}}
      {\prod_{1 \le j < k \le n}(y_{j}-y_{k}+i)},
\nonumber 
\end{eqnarray}
where functions $\mathfrak{a}(v)$ and $G(v,\xi)$ are solutions of the NLIE: 
\begin{eqnarray}
\log \mathfrak{a}(v)=-\frac{h}{T}+\frac{2J}{v(v+i)T} - \int_{C}\frac{dy}{\pi}
 \frac{\log (1+\mathfrak{a}(y))}{1+(v-y)^{2}},
  \label{nlie1}
\end{eqnarray}
\begin{eqnarray}
G(v,\xi)&=& -\frac{1}{(v-\xi )(v-\xi-i)} \nonumber \\ 
&& + \int_{C}\frac{dy}{\pi}
 \frac{1}{1+(v-y)^{2}} \frac{G(y,\xi)}{1+\mathfrak{a}(y)}.
 \label{nlie2}
\end{eqnarray}
Here the contour $C$ surrounds the real axis anti-clockwise manner. 

First, let us calculate the HTE of $\mathfrak{a}(v)$ from the NLIE (\ref{nlie1}). 
This is done by a similar procedure in \cite{DD92}, where a certain order of the HTE for 
the free-energy was calculated from a NLIE. 
We assume the following expansion for small $J/T$,
\begin{equation}
\mathfrak{a}(v)=\exp\left(\sum_{k=1}^{\infty}a_{k}(v)\left( \frac{J}{T} \right)^{k}\right).
\label{a-ex}
\end{equation}
Substituting (\ref{a-ex}) into (\ref{nlie1}), and 
comparing coefficients of $(J/T)^m$ on both sides, 
we obtain an integral equation over $\{ a_{k}(v) \}_{k=1}^{m}$ 
for each $m$ ($m \in {\mathbb Z}_{\ge 1}$). 
As the resultant integral equation is linear with respect to $a_{m}(v)$, 
we can solve it recursively. For example, we obtain
\begin{eqnarray}
&& a_{1}(v)=-\frac{h}{J}-\frac{2 i}{v(1 + v^2)},
\nonumber  \\ 
&& 
a_{2}(v)= \frac{h}{J( 1 + v^2 )} + 
    \frac{2 i v}{( 1 + v^2 )^2}, 
\nonumber  \\ 
&&
a_{3}(v)= -\frac{h}{J(1+v^2)}.
\end{eqnarray}
Note that only $a_{1}(v)$ has a pole at the origin.  
Next let us consider the integral equation (\ref{nlie2}). We assume the following 
expansion for small $J/T$,
\begin{equation}
G(v,\xi)=\sum_{k=0}^{\infty}g_{k}(v,\xi)\left(\frac{J}{T}\right)^{k}.
 \label{ex-g}
\end{equation}
In a similar way, we can determine the coefficients $g_{k}(v,\xi)$ 
successively by using the results on (\ref{nlie1}). For example, we 
obtain
\begin{eqnarray}
&& g_{0}(v,\xi)=\frac{-i}{( 1 + {( v - \xi ) }^2 ) ( v - \xi ) },
\nonumber \\
&& g_{1}(v,\xi)= \frac{i(2v -\xi)}
  {( 1 + v^2 ) ( 1 + {( v - \xi ) }^2 ) 
    ( 1 + {\xi}^2 )}
 \nonumber \\
&& \hspace{10pt} +\frac{h}{2J(1+(v-\xi)^2)},
 \nonumber \\
&& g_{2}(v,\xi)= -\frac{i(2v -\xi)\xi^2}
  {( 1 + v^2 ) ( 1 + {( v - \xi ) }^2 ) 
    ( 1 + {\xi}^2 )^{2} }
 \nonumber \\
&& \hspace{10pt} -
\frac{h(2 + 2 v^2 - 2 v \xi + {\xi}^2)}
  {2J( 1 + v^2 ) ( 1 + {( v - \xi ) }^2 ) 
    ( 1 + {\xi}^2 ) }.
\end{eqnarray}
Note that only $g_{0}(v,\xi)$ has a pole at $v=\xi$. 
Finally, substituting (\ref{a-ex}) and (\ref{ex-g}) into (\ref{efp}), 
we can obtain the HTE of $P(n)$. Unexpectedly we have found that 
we only have to 
calculate residues at the origin. In fact, in this way, we could calculate 
the HTE of the $P(n)$ for small ${n}$ ($n \in \{2,3,4,5,6\}$). The result up 
to the order $O((J/T)^4)$ is compactly presented as 
\begin{eqnarray}
P(n)&=&\frac{1}{2^n}+\frac{-2J( -1 + n )  + hn}{2^{1 + n} T}
\nonumber \\ 
&+& \Bigl\{ 4J^2( -4 + n ) ( -1 + n )  + 
    h^2( -1 + n ) n  
\nonumber \\ 
 && -4hJ( 2 + ( -1 + n ) n ) \Bigr\} \frac{1}{2^{3 + n}T^2}
\nonumber \\    
&+& \Bigl\{ 12hJ^2{( -2 + n ) }^2
       ( -1 + n )  + h^3( -3 + n ) n^2 
\nonumber \\ 
&& - 8J^3( -24 + ( -9 + n ) ( -3 + n ) n
         ) 
\nonumber \\ 
&& - 6h^2J( -1 + n ) 
       ( 2 + ( -1 + n ) n )  \Bigr\} 
 \frac{1}{3 \cdot 2^{4 + n}T^3}
\nonumber \\
&+&  \Bigl\{ 24h^2J^2{( -2 + n ) }^2
       {( -1 + n ) }^2 
\nonumber \\ 
&& + 16J^4( -5 + n ) 
       ( -32 + n( 26 + ( -17 + n ) n )  ) 
\nonumber \\
&& + h^4n( 2 + n( 3 + ( -6 + n ) n ) 
         )  
\nonumber \\
&& - 32hJ^3( 24 + 
         ( -9 + n ) n( 6 + ( -3 + n ) n ) 
         ) 
\nonumber \\
&& - 8h^3J( -4 + 
         ( -4 + n ) n( 3 + n^2 )  )  \Bigr\} 
        \frac{1}{3 \cdot 2^{7 + n}T^4}
\nonumber \\ 
&+&   O\left(\left(J/T\right)^5\right). \label{pn-general}
\end{eqnarray}
We observe that $P(n)$ has the following form
\begin{eqnarray}
P(n)&=&\frac{1}{2^n}\sum_{m=0}^{\infty}\frac{1}{m!}
 \left( \sum_{k=0}^{m} p_{m,k}(n) \left(\frac{h}{J}\right)^{k} \right) \left(\frac{J}{2T}\right)^m, \nonumber 
\end{eqnarray}
where $p_{0,0}(n)=1$ and $p_{m,k}(n)$ for $m \in {\mathbb Z}_{\ge 1}$ 
are functions of $n$, which are independent of $J$, $h$ and $T$. 
If we admit that $p_{m,k}(n)$ is a polynomial of $n$ whose degree is at most $m$, 
our formula (\ref{pn-general}) is also valid for any $n \in {\mathbb Z}_{\ge 2}$ as 
the $m$-th order polynomial is determined by distinct $m+1$ points. 

Next we fix the magnetic field ${h}$ to certain values and calculate the HTE to much 
higher order. Then we have succeeded in obtaining coefficients of $P(3)$ up to 
the order ${O((J/T)^{42})}$ in the case of ${h=0}$. For a finite ${h}$,  we can  
calculate them at least up to the order ${O((J/T)^{30})}$. 
It will not be easy to obtain HTE coefficients, in particular under the magnetic field,
to such a high order by other method except for the free models. 
We list some of our results on $h=0$ case in Table \ref{table1}. 

\begin{figure}[htbp]
\includegraphics[width=0.36 \textwidth]{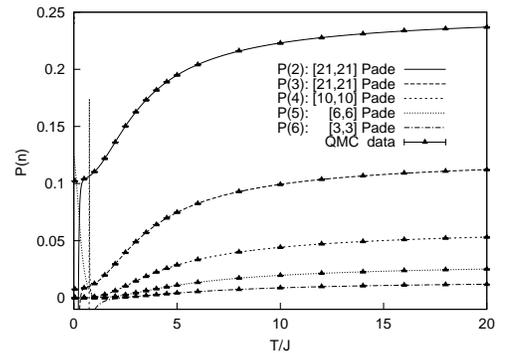} 
\caption{\label{fig:1}  $P(n)$ for $J>0$ at ${h=0}$.} 
\end{figure}
\begin{figure}[htbp]
\includegraphics[width=0.36 \textwidth]{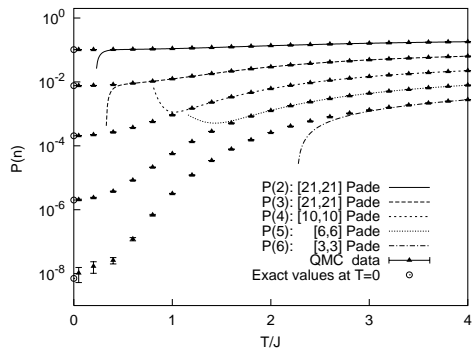} 
\caption{\label{fig:2} 
 $P(n)$ for $J>0$ at ${h=0}$ in low ${T}$. The exact values at ${T=0}$ were obtained in 
\cite{BK01,BKNS02,BKS04}.  
} 
\end{figure}
\begin{figure}[htbp]
\includegraphics[width=0.36 \textwidth]{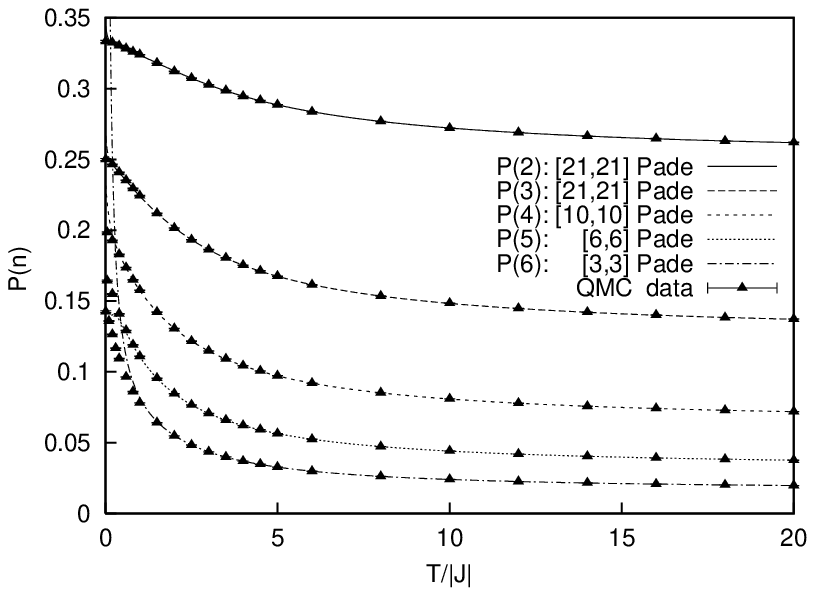} 
\caption{\label{fig:3} $P(n)$ for $J<0$ at ${h=0}$.} 
\end{figure}
\begin{figure}[htbp]
\includegraphics[width=0.36 \textwidth]{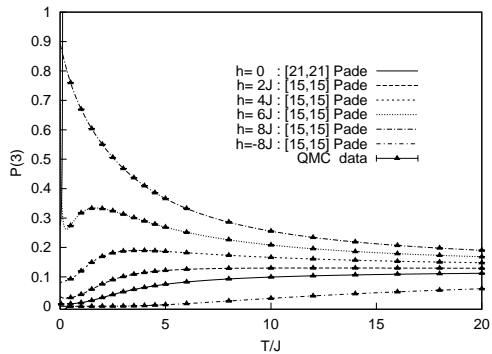} 
\caption{\label{fig:4} $P(3)$ for $J>0$ at ${h=0,2J,4J,6J,\pm8J}$.} 
\end{figure}
\begin{figure}[htbp]
\includegraphics[width=0.36 \textwidth]{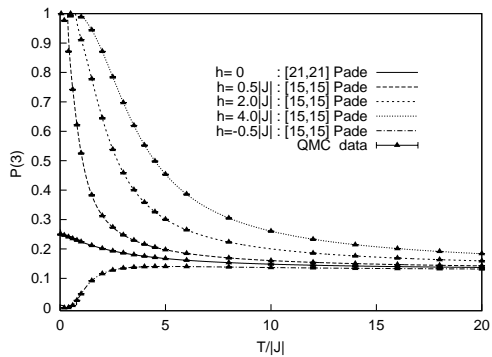} 
\caption{\label{fig:5} $P(3)$ for $J<0$ at ${h=0,\pm 0.5|J|,2|J|,4|J|}$.} 
\end{figure}
Moreover we have applied the Pad\'{e} approximation to our HTE and plot the results 
in Fig.~1-Fig.~5. Note that although the formula (\ref{efp}) was originally derived 
for $J>0$ case, our HTE results can be analytically continued to $J<0$ case.
For comparison, we have also performed Quantum Monte Carlo simulation (QMC) 
using recent open source softwares in ALPS project \cite{ALPS}. Especially we have 
chosen the SSE algorithm \cite{SSE} so as to treat finite magnetic field cases. We have 
performed the simulations with the system size ${L=128}$.
In Fig.~1-Fig.~5, these QMC data are represented by solid 
triangles, which show excellent agreements with the HTE results. Discrepancy appears 
only in the very low temperature regions, 
where even the Pad\'{e} approximation of the HTE  
ceases to converge. We omit these apparent deviations of the Pad\'{e} approximation 
in Fig.~2. We remark that we have also tested the validity of our  general formula 
(\ref{pn-general}) up to ${n=20}$.

In the case of ${J>0}$ and ${h=0}$, we see that ${P(n)}$ monotonously 
increases as the temperature increases. On the other hand, it decreases monotonously 
for ${J<0}$. In this case we have found ${P(n) \to 1/(n+1)}$ as ${T \to 0}$. 
Another interesting observation in Fig.~\ref{fig:4} is that, when ${J>0}$, a peak appears 
for positive values of the magnetic field. Its position moves from $T= \infty $ to $0$ as $h$ increases. 
For example, the peak position $T^{\rm max}$ and the peak $P(3)^{\rm max}$  are  given by 
$(h/J,T^{\rm max}/J,P(3)^{\rm max})=(2, 12.030, 0.13015)$, $(4, 3.7467, 0.18973)$,
 $(6, 1.6904, 0.33416)$, respectively. Note that in this case, the critical field is 
${h_{\rm c}=8J}$ at ${T=0}$, where all the spins are directed upward.  

In conclusion we have found that our HTE method is very powerful to evaluate the integral 
formula for ${P(n)}$ at finite temperature. As an alternative way, it may be possible to 
solve the NLIEs (\ref{nlie1}) and (\ref{nlie2}) numerically and perform numerical 
integration for the multiple integrals in (\ref{efp}). We have tried it, but found it 
hard to get reliable numerical results even for $P(3)$. 

It is straight forward to generalize the results in this letter to more general correlation 
functions.  Actually as for the nearest and the next-nearest-neighbor correlation functions 
for $h=0$, we can immediately calculate their HTEs from our results through the relations 
$\langle S^{z}_{j}S^{z}_{j+1} \rangle =P(2)-1/2$, 
$\langle S^{z}_{j}S^{z}_{j+2} \rangle =2(P(3)-P(2)+1/8)$,   
from which we can obtain the coefficients whose order is higher than the ones by other method \cite{F03}. 
We can evaluate the HTE for other complicated correlation functions  based on 
the multiple integral formula on the density matrix of the $XXZ$ chain at finite 
temperatures \cite{GKS04-3}. We will report on details in the forthcoming paper \cite{STT05}. 

The authors are grateful to F.~G\"{o}hmann, A.~Kl\"{u}mper, K.~Sakai, and M.~Takahashi for valuable 
discussions. Special thanks are due to M.~Inoue for private communication on HTE on (\ref{nlie1}) and  
T.~Kato for an introduction of Quantum Monte Carlo simulations. 
ZT is supported by Grant-in-Aid for Scientific Research 
from JSPS (no. 16914018). MS is  supported by Grant-in-Aid for 
Young Scientists  (no. 14740228).
\
\begingroup
\squeezetable
\renewcommand{\arraystretch}{1.8}
\begin{table*}
\caption{\label{table1}Series coefficients $p_{k}(n)$ for 
the high temperature expansion of 
$P(n)=\sum_{k}p_{k}(n) (\frac{J}{T})^{k}$ at ${h=0}$.}
\begin{ruledtabular}
\begin{tabular}{llll|ll|ll} 
$k$ & $p_{k}(3)$ & $21$ & $- \frac{475666635106757}{89391802500}$ & $ k$ &
 $p_{k}(4)$ & $k$ & $p_{k}(5)$ \\
 $0$ &  $\frac{1}{8}$ & $22$ & $\frac{73241259005444676467}{204177262038750}$ & $0$ & $\frac{1}{16}$
 & $0$ & $\frac{1}{32} $ \\
 $1$ & $- \frac{1}{4}$ & $23$ & $\frac{14342832948901512027127}{39447047025886500}$ 
& $1$ & $- \frac{3}{16}$ & $ 1$ & $- \frac{1}{8}   $ \\
 $2$ & $- \frac{1}{8}$ & $24$ & $- \frac{354494436182818781071}{297489042427500}$ & $2$ & $0$ & $2$ & 
$\frac{1}{16} $ \\
 $3$ & $\frac{1}{2}$ & $25$ & $- \frac{39877735294663490409548941}{14792642634707437500}$ & $3$
 & $\frac{11}{24}$ & $3$ & $\frac{1}{3} $ \\
 $4$ & $\frac{5}{6}$ & $26$ & $\frac{491132965711734931876859809}{192304354251196687500}$ &
 $4$ & $\frac{17}{48}$ & $4$ & $0 $ \\
 $5$ & $- \frac{21}{20}$ & $27$ & $\frac{66961305287544998239794361}{4767876551682562500}$
 & $5$ & $- \frac{343}{240}$ & $5$ & $- \frac{21}{16}   $ \\
 $6$ & $- \frac{487}{120}$ & $28$ &
 $\frac{9692024436454844294876678309}{4038391439275130437500}$ & 
$6$ & $- \frac{937}{360}$ & $6$ & $- \frac{301}{288}$ \\
 $ 7$ & $\frac{271}{630}$ & $29$ & 
$- \frac{4098815896973029894033624285217}{70268011043387269612500}$ & $7$ 
& $\frac{221}{63}$ & $7$ & $\frac{2923}{630} $ \\
 $ 8$ & $\frac{5161}{315}$ & $30$
 & $- \frac{232435776187690664677091074186001}{3513400552169363480625000}$ & $8$ 
& $\frac{17267}{1260}$ & $8$ & $
   \frac{84319}{10080} $ \\
 $ 9$ & $\frac{1105}{84}$ & $31$ 
& $\frac{46326083992727076268552170704552473}{245059688513813102773593750}$
 & $9$ & $- \frac{185}{81}$ & $9$ & $
   - \frac{566639}{45360}   $ \\
 $ 10$ & $- \frac{256276}{4725}$ & $32$
 & $\frac{8813049514657368316161218121220189}{18850745270293315597968750}$ & $10$ & $
   - \frac{668573}{11340}$ & $10$ & $- \frac{47129}{1008}   $ \\
$11$ & $- \frac{21532949}{207900}$ & $33$ &
 $- \frac{23560005035480782300479589437217351}{63427213497692803070812500}$ & $11$ & 
$- \frac{456671}{10395}$ & $11$ & $\frac{324749}{27720} $ \\
 $ 12$ & $\frac{420091}{3300}$ & $34$ & $
   - \frac{27945437643781625566445986974785072}{11720245972399757089171875}$ & $12$
 & $\frac{4924421}{23760}$ & $12$ & $\frac{3162705559}{14968800} $ \\ 
\cline{7-8}
 $13$ & $\frac{2660539279}{4864860}$ & $35$ & 
$- \frac{926901382890398878135256848861535464573}{1374784852562491506559860937500}$ & $ 13$
 & $\frac{12185848483}{32432400}$ & $k$ & $p_{k}(6) $  \\
 $ 14$ & $- \frac{6793954613}{340540200}$ & $36$
 & $\frac{18696605034197142286345924338927419869073}{1924698793587488109183805312500}$ & $14$ & $
   - \frac{180113062933}{340540200}$ & $0$ & $\frac{1}{64} $ \\
 $15$ & $- \frac{37743598006}{16372125}$ & $37$ & 
$\frac{4364774409085808451234535985560619072990519}{356069276813685300199003982812500}$ & $15$ & $
   - \frac{38034664397}{18243225} $ & $1$ & $- \frac{5}{64}   $ \\
 $16$ & $- \frac{1327276364741}{638512875}$ & $ 38$ &
 $- \frac{71422507032359703142873531724137016664174319}{2341840243659237935924218502343750} $ & $16$ & $
   \frac{29988604741}{98232750}$ & $2$ & $\frac{5}{64} $ \\
 $ 17$ & $\frac{336925562547463}{43418875500}$ & $39$ & $
   - \frac{7272089125489543224633329075851883666829606991}{87949111372980269149153983754687500}$ & 
$ 17$ & $\frac{11380401164189}{1240539300}$ &
 $3$ & $\frac{13}{64} $ \\
 $18$ & $\frac{26889108889501}{1669956750}$ & $40$ & 
$\frac{634149726978610912172163815072882174172080282077}{11873130035352336335135787806882812500}$ &
 $18$ & $\frac{1902555753434863}{260513253000}$ & $4$ & $- \frac{17}{96}   $ \\
 $19$ & $- \frac{43793345212356097}{2474875903500}$ & $ 41$
 & $\frac{66831923892050783799016502839247559821880338345693}{162266110483148596580189100027398437500}$
 & $19$ & $- \frac{481157848754889239}{14849255421000}$ & $ 5$ & $- \frac{19}{20}$ \\
 $20$ & $- \frac{3162447776015376619}{37123138552500}$ & $ 42$ & $
   \frac{556085532621818295062304191328067921669214230014169}{3407588320146120528183971100575367187500}$
 & $20$ & $- \frac{139195329414966479}{2249887185000}$ & $6$ & $ \frac{31}{480} $
\end{tabular}
\end{ruledtabular}
\end{table*}

\endgroup

\noindent
\footnotesize{E-mail address (Z. Tsuboi): tsuboi@issp.u-tokyo.ac.jp} \\
\footnotesize{E-mail address (M. Shiroishi): siroisi@issp.u-tokyo.ac.jp}

\end{document}